\pdfoutput=1
\documentclass[twocolumn,english,aps,prb,groupedaddress]{revtex4}
\usepackage[T1]{fontenc}
\usepackage[latin9]{inputenc}
\usepackage[letterpaper]{geometry}
\usepackage{array}
\usepackage{graphicx}
\usepackage{amssymb}

\makeatletter

\providecommand{\tabularnewline}{\\}

\@ifundefined{textcolor}{}
{%
 \definecolor{BLACK}{gray}{0}
 \definecolor{WHITE}{gray}{1}
 \definecolor{RED}{rgb}{1,0,0}
 \definecolor{GREEN}{rgb}{0,1,0}
 \definecolor{BLUE}{rgb}{0,0,1}
 \definecolor{CYAN}{cmyk}{1,0,0,0}
 \definecolor{MAGENTA}{cmyk}{0,1,0,0}
 \definecolor{YELLOW}{cmyk}{0,0,1,0}
 }

\@ifundefined{definecolor}
 {\usepackage{color}}{}
\usepackage{epstopdf}\usepackage{epsfig}

\makeatother

\usepackage{babel}

\begin{document}

\title{Fabrication of stable and reproducible sub-micron tunnel junctions}

\author{I. M. Pop, T. Fournier, T. Crozes, F. Lecocq, I. Matei, B. Pannetier,
O. Buisson and W. Guichard}

\affiliation{$^{1}$Institut Néel, CNRS-UJF, 25 Rue des Martyrs, 38042 Grenoble,
France}

\date{\today}
\begin{abstract}
We have performed a detailed study of the time stability and reproducibility
of sub-micron $Al/AlO_{x}/Al$ tunnel junctions, fabricated using
standard double angle shadow evaporations. We have found that by aggressively
cleaning the substrate before the evaporations, thus preventing any
contamination of the junction, we obtained perfectly stable oxide
barriers. We also present measurements on large ensembles of junctions
which prove the reproducibility of the fabrication process. The measured
tunnel resistance variance in large ensembles of identically fabricated
junctions is in the range of only a few percents. Finally, we have
studied the effect of different thermal treatments on the junction
barrier. This is especially important for multiple step fabrication
processes which imply annealing the junction.
\end{abstract}
\maketitle
Sub-micron sized metal$/AlO_{x}/$metal tunnel junctions are used
in a wide range of applications in many fields, from superconducting
and single charge electronics to calorimetry, nanomagnetism and spintronics.
However, the time instability of the $AlO_{x}$ barrier is a frequently
reported drawback. Systematic measurements of junction aging showed
that the resistance of the tunnel barrier could double its value within
a period of days \cite{Koppinen2007}. Similarly, a reduction of the
capacitance $C$ of the junctions was also observed \cite{Nesbitt2007}.
The junction aging is usually associated with two types of phenomena.
Either $(i)$ the diffusion of oxygen atoms from the oxide barrier
to the electrodes \cite{SHIOTA1992} or $(ii)$ the change of the
chemical composition of the barrier, by absorption or desorption of
atoms or molecules other than oxygen \cite{GATES1984}. It has been
shown that mechanism $(i)$ plays a secondary role in the aging of
the barrier, only accounting for the slow drift of the junction parameters
at long time scales. Moreover, this slow diffusion of oxygen atoms
can be suppressed by surface nitridation of the electrodes \cite{SHIOTA1992a}.
Mechanism $(ii)$ is believed to play the dominant role in the aging
of the junction barrier. Vacuum anneals at temperatures between $200\,^{o}C$
and $450\,^{o}C$ will accelerate the relaxation processes in the
$AlO_{x}$ barrier and the resulting junctions show improved characteristics
\cite{Koppinen2007,Julin2010}.

The origin of the junction contamination, which leads to the chemical
relaxation in mechanism $(ii)$, has been up to now subject of speculation.
Using a standard bilayer double angle evaporation technique\cite{DOLAN1977},
Koppinen et al. \cite{Koppinen2007} observed a decrease in the junction
aging when the substrate was cleaned with an oxygen plasma just before
the electrodes evaporation. This observation suggests that the aging
of the barrier is linked to the existence of resist residues in the
vicinity of the junction during the deposition. Indeed, by using fabrication
methods which avoid the direct contact between the photoresist and
the insulating layer, it is possible to obtain perfectly stable junctions
\cite{GURVITCH1983,MOROHASHI1991}. In the following we will show
that by aggressively cleaning the substrate before the $Al$ deposition,
we could obtain perfectly stable $Al/AlO_{x}/Al$ junctions, using
the standard bilayer double angle evaporation technique. Further on,
we will show evidence that the junction aging is a consequence of
the oxide barrier contamination from resist residues.

We fabricate the samples using the standard \textit{Dolan bridge}
technique\cite{DOLAN1977}. We produce the suspended resist mask using
a PMMA/PMMA-MAA bilayer, deposited on a $Si/SiO_{2}$ substrate. The
circuit patterns are written using e-beam lithography. After development
we perform two shadow evaporations of $Al$ in a UHV system with a
base pressure of $10^{-9}\mbox{mBar}$. In order to obtain tunnel
junctions, between the two evaporations we oxidize the first $Al$
layer in a pure oxygen atmosphere at different pressures $P_{oxydation}$,
depending on the sample (see Table \ref{Flo:TableJunctionStabilityInTime}). 

\begin{figure}
\begin{centering}
\includegraphics[width=0.9\columnwidth]{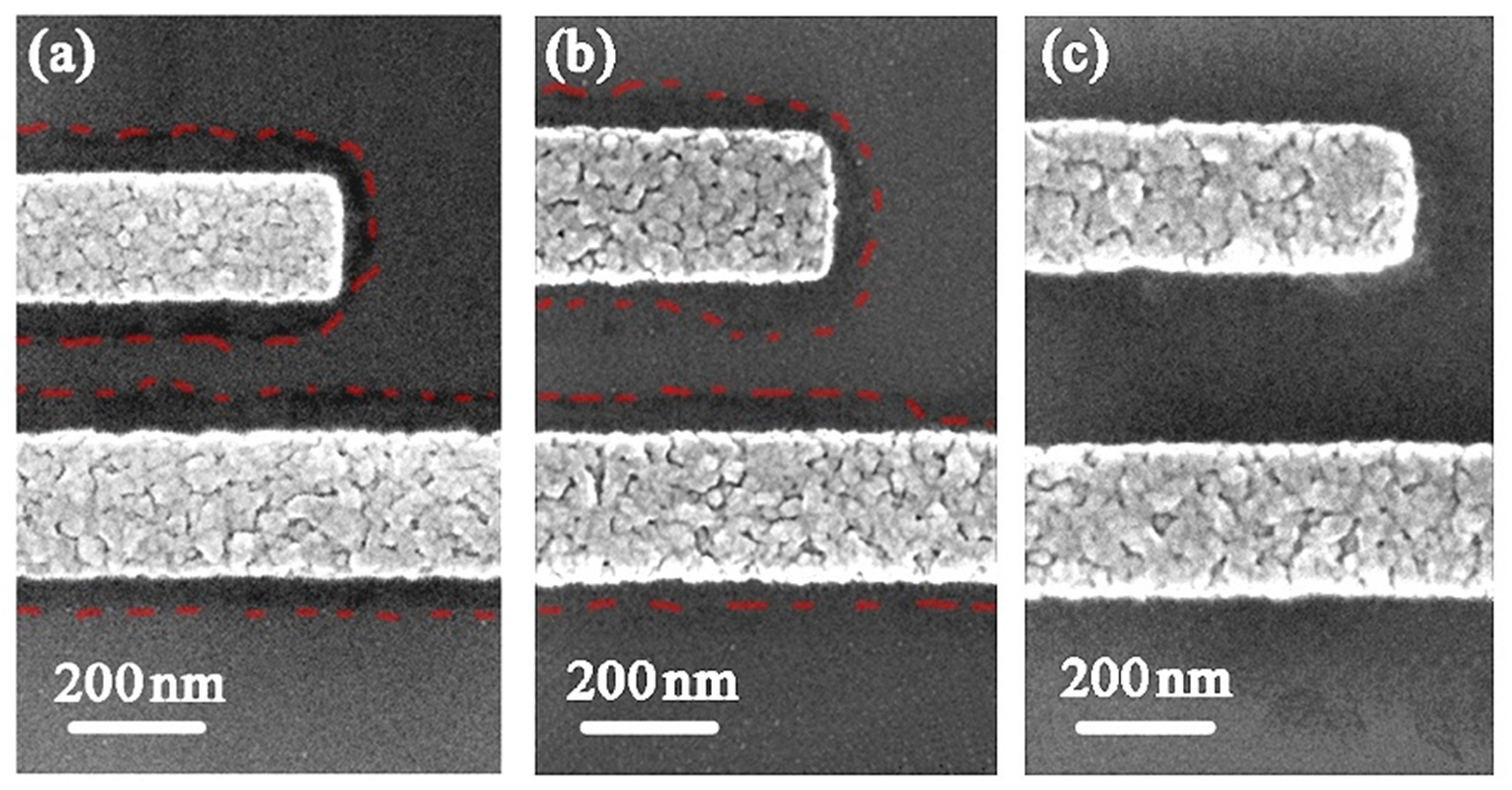}
\par\end{centering}

\caption{SEM images of zero angle deposited $Al$ wires, for three different
development sequences. The images are taken at high contrast in order
to observe the resist residues, which are visible as black traces.
For clarity, we highlighted them with the red dotted line. (a) Standard
MIBK/IPA 1/3 (30s) and IPA (60s) development. (b) After the standard
development we added an Ethanol/IPA 1/3 (60s) step. We notice that
the wafer is significantly cleaner, even though some resist traces
can still be seen. (c) After the standard development we performed
an oxygen plasma RIE. Notice that the surface of the wafer is now
completely resist free.}

\label{Flo:DifferentCleaningOfSurface}
\end{figure}

In Fig. \ref{Flo:DifferentCleaningOfSurface}a we show a SEM image
of Al wires deposited after a standard development of the resist bilayer.
 In the vicinity of the electrodes we see a significant contamination
with residual resist. In order to clean these resist residues, we
have tried two processing steps. First, after the standard development,
we plunge the wafer for 1 minute in a mixture of Ethanol and IPA.
Ethanol is an aggressive solvent and it removes large volumes of resist
from the undercut. In Fig. \ref{Flo:DifferentCleaningOfSurface}b
we show a SEM image of the resulting Al wires. We notice that we have
removed part of the residues, but the substrate is not yet completely
clean. The second wafer cleaning step that we tried is the Reactive
Ion Etching (RIE) in oxygen plasma. The plasma parameters were adapted
to ensure a slow, isotropic etching\cite{RIEparamGrenoble}. This
is not obvious because we have to clean a relatively large undercut
surface, compared to the small access opening in the top resist. The
SEM image of the resulting $Al$ wires, presented in Fig. \ref{Flo:DifferentCleaningOfSurface}c,
shows that the oxygen plasma RIE has completely cleaned the resist
residues. 

Using the oxygen plasma RIE cleaning step, we have fabricated several
sets of $\sim100$ junctions that we have contacted and measured individually
at room temperature. %
\begin{figure}[tbph]
\medskip{}

\begin{centering}
\includegraphics[width=0.8\columnwidth]{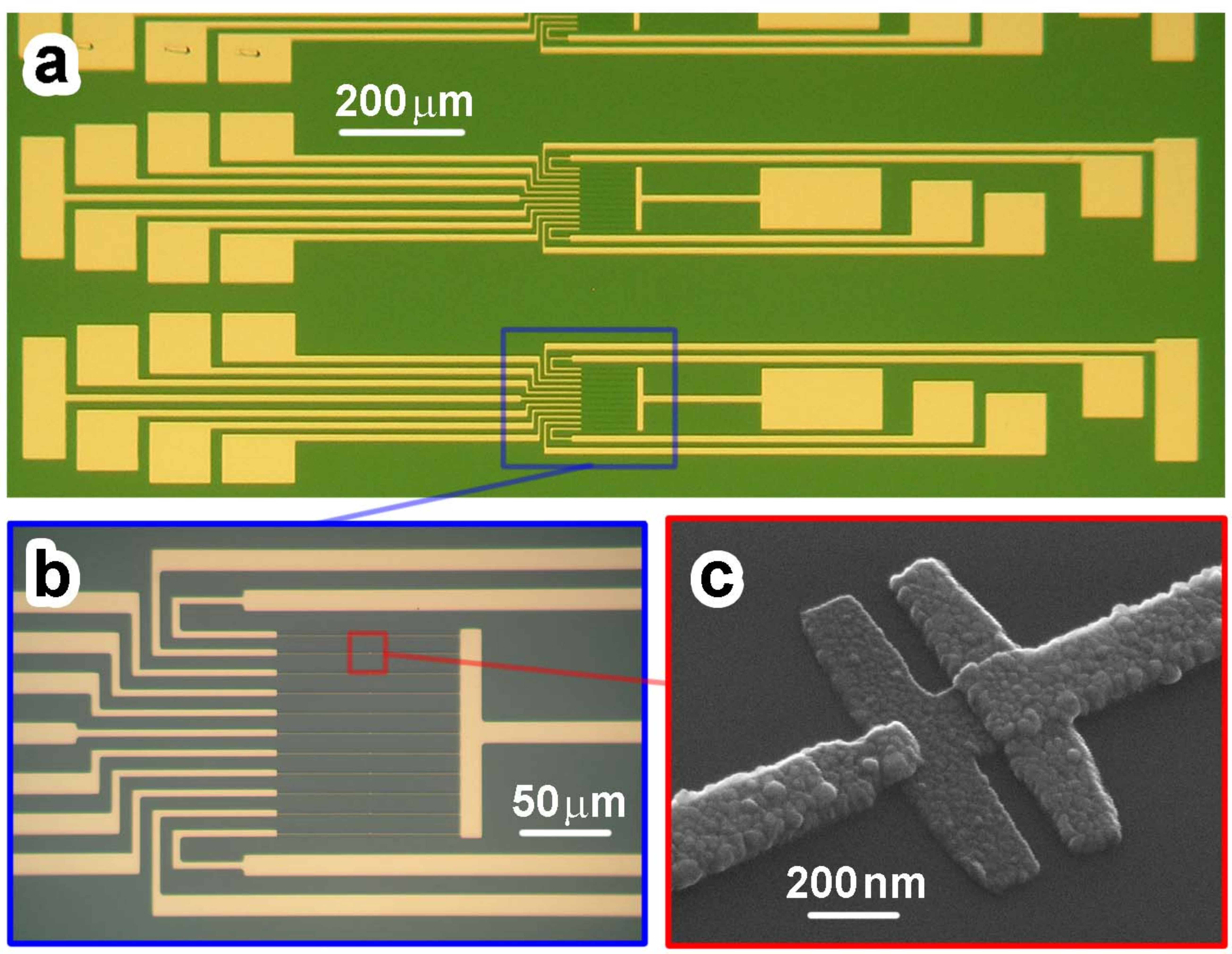}\bigskip{}

\par\end{centering}

\begin{centering}
\includegraphics[width=0.6\columnwidth]{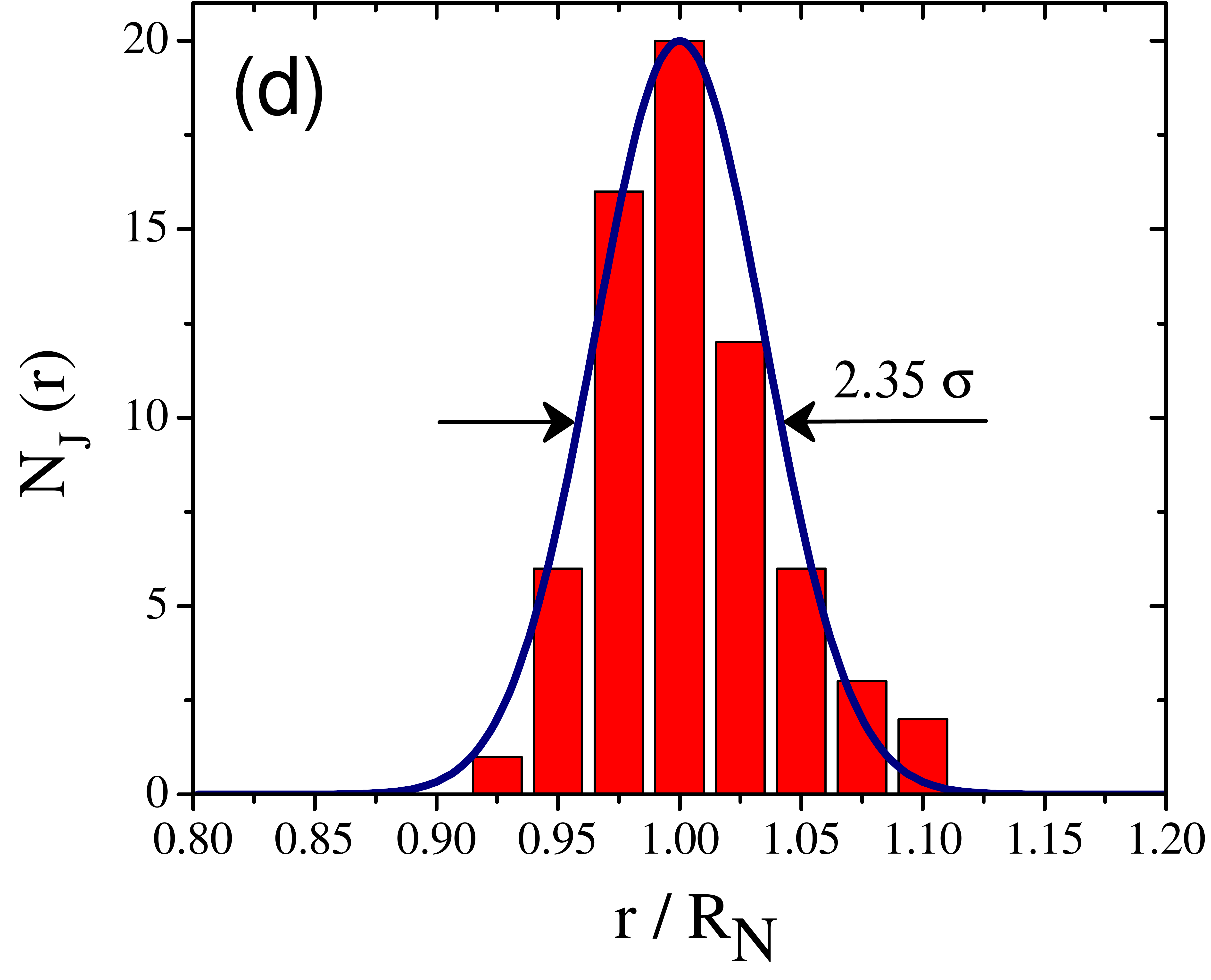}
\par\end{centering}

\caption{Images of the circuit fabricated to measure the time stability and
the variance of the junction's tunnel barrier. In (a) and (b) we show
optical images of one of the junction sets at different magnifications.
(c) SEM image of one junction. (d) Measured histogram of the tunnel
barrier resistance, for identically fabricated junctions with a surface
of $0.03\,\mbox{\ensuremath{\mu m^{2}}}$. We also represented the
width at half height for the Gaussian fit (blue line), which is $\sim2.35\cdot\sigma$,
where $\sigma$ is the variance. For clarity, the $x-axis$ is reported
in units of the mean value for the resistance: $R_{N}$.}

\label{Flo:SamplePhoto}
\end{figure}
 The geometry of the sample (Fig. \ref{Flo:SamplePhoto}) was optimized
to enable the packing of a large number of identical junctions in
a small area of the wafer, in order to minimize the variations of
resist thickness. The surface $S$ of the junctions depends on the
sample, ranging from $0.02\,\mbox{\ensuremath{\mu m^{2}}}$ to $0.2\,\mbox{\ensuremath{\mu m^{2}}}$.
At low temperatures, the junctions have critical current densities
in the range of $\sim1\,\mu\mbox{A/\ensuremath{\mu m^{2}}}$. This
class of $Al/AlO_{x}/Al$ junctions is widely used in superconducting
quantum electronics \cite{Buisson2009}.

In Fig. \ref{Flo:SamplePhoto}d we present a typical histogram for
the tunnel resistance $r$ measured at room temperature. The mean
value of the resitance $\left\langle r\right\rangle =R_{N}$, depending
on the sample, is of the order of a few $\mbox{k\ensuremath{\Omega}}$.
The distribution is well fitted (see the blue line in Fig. \ref{Flo:SamplePhoto}d)
by a Gaussian law: $N_{J}(r)=N_{tot\,}\left(1/\sqrt{2\pi\sigma^{2}}\right)exp\left[-\left(r/R_{N}-1\right)^{2}/2\sigma^{2}\right].$
Here $\sigma^{2}$ is the variance, in units of $R_{N}^{2}$ and $N_{tot}$
is the total number of measured junctions. The measurements in Fig.
\ref{Flo:SamplePhoto}d can be well fitted by taking the value $\sigma=3.5\mbox{\%}$.
All measured junction ensembles showed similar variance values, within
a few percents.

We measured the average tunnel resistance of the junctions at room
temperature, directly after the fabrication process $R_{N}\left(0\right)$
and during the following 4 weeks $R_{N}\left(t\right)$. We studied
several ensembles of $\sim100$ junctions (see Fig. \ref{Flo:SamplePhoto})
for which the fabrication conditions and the storage were identical.
The results of our observations are presented in Table \ref{Flo:TableJunctionStabilityInTime}.
\begin{table}[tbph]
\begin{centering}
\begin{tabular}{|>{\centering}m{0.1\columnwidth}|>{\centering}m{0.25\columnwidth}|>{\centering}m{0.13\columnwidth}|>{\centering}m{0.2\columnwidth}|>{\centering}m{0.22\columnwidth}|}
\hline 
\textbf{chip } & $P_{oxydation}$ & storage & \textbf{$\begin{array}{c}
R_{N}\left(0\right)\times S\\
{}[\Omega\times\mu m^{2}]\end{array}$} & 

\textbf{$\frac{R_{N}\left(t\right)-R_{N}\left(0\right)}{R_{N}\left(0\right)}$ }\tabularnewline
\hline
\hline 
$A$ & $5\cdot10^{-2}\, mbar$ & vacuum & 210 & $0\%$\tabularnewline
\hline 
$B$ & $4\cdot10^{-2}\, mbar$ & atm & 180 & $0\%$\tabularnewline
\hline 
$C$ & $5\cdot10^{-3}\, mbar$ & atm & 160 & $+10\%$\tabularnewline
\hline
\end{tabular}
\par\end{centering}

\caption{Measurement of junction stability, during 4 weeks after fabrication.
The error bar for the measurement is $\sim1\%$. The oxidation time
for all samples is 5 minutes. }

\label{Flo:TableJunctionStabilityInTime}
\end{table}
\begin{figure}[tbph]
\begin{centering}
\includegraphics[width=0.99\columnwidth]{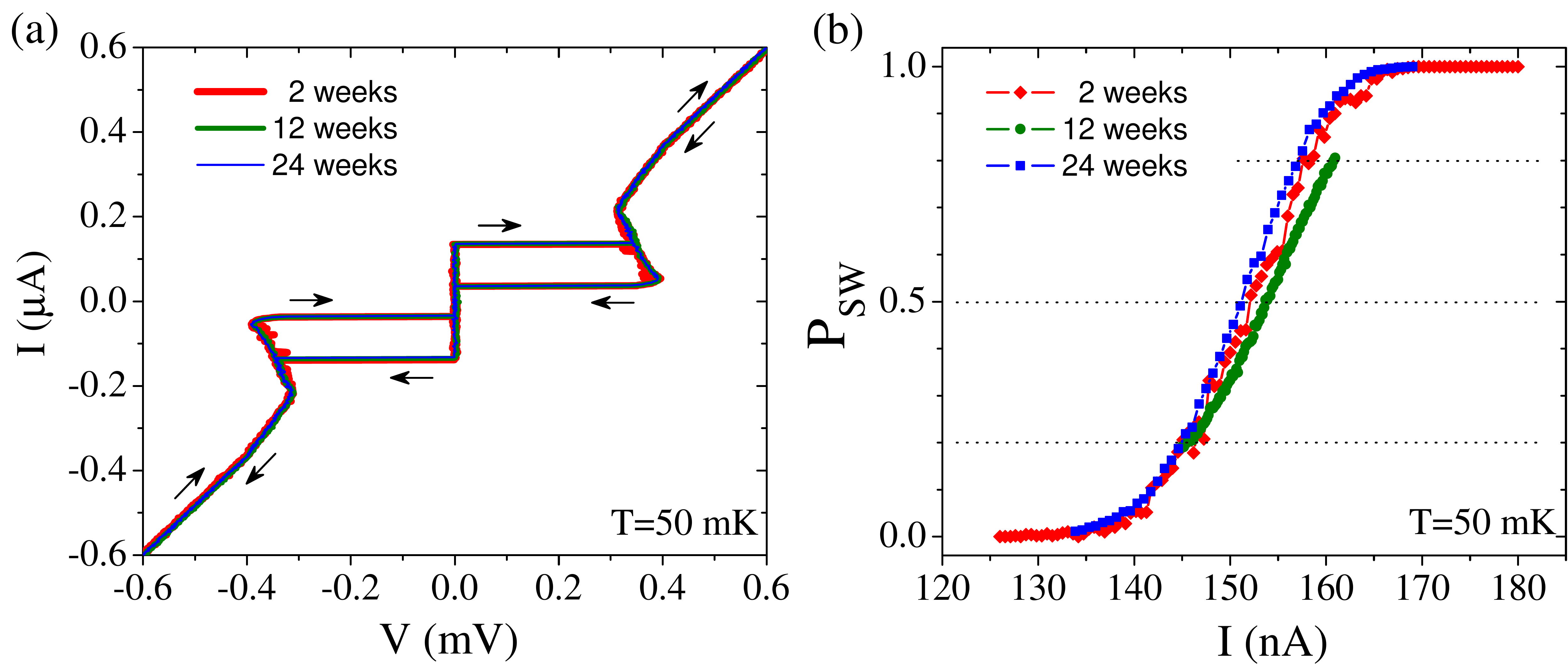}
\par\end{centering}

\caption{Current-voltage curves (in a) and switching current measurements (in
b) for the same junction from group A, measured 2 weeks (in red),
12 weeks (in green) and 24 weeks (in blue) after the fabrication.
Between the measurements, the junction was stored at room temperature
and atmospheric pressure.}

\label{Flo:IVcurvesDifferentDates}
\end{figure}
 For the junctions of chip $A$ and $B$ we did not observe any aging.
For sample $C$ we measured a slight increase in the tunnel barrier,
which we attribute to the adsorption of oxygen from the surrounding
atmosphere. This adsorption is favored by the high reactivity of the
barrier, which was very weakly oxidized. All the measured tunnel barriers
are practically stable if compared to the previously reported increases
of $100\mbox{\%}$ up to $300\mbox{\%}$ within the first weeks after
the fabrication\cite{Koppinen2007,Nesbitt2007}. 

One of the junctions from group $A$ was measured at $T=50\,\mbox{mK}$
in the superconducting regime, 2 weeks, 12 weeks and 24 weeks after
the fabrication. The measured $IV$ characteristics and the switching
curves are plotted in Fig. \ref{Flo:IVcurvesDifferentDates}. Notice
that the results of the 3 measurements at $T=50\,\mbox{mK}$ are practically
identical. We did not observe any aging of the tunnel barrier, the
critical current or the retrapping current. The switching current
and the width of the switching curves, both depend on the junction
capacitance $C$, and show no aging between the three measurements.
Thus, we conclude that the capacitance of the junction is also completely
stabilized. From the small difference of $\pm5\mbox{nA}$between the
three switching curves, we infer the error bar of our measurement,
which is smaller than $5\%$.

In the following, we study the effects of annealing on the oxide tunnel
barrier. We have measured the effect of vacuum thermal cycles on the
tunnel barrier of our junctions. The steps of one thermal cycle are
the following: $10$ minutes heating, $10$ minutes regulating the
temperature to a constant value and $20$ minutes cooling down. In
Ref. \cite{Koppinen2007} it has been shown that the resistance of
samples annealed at $200\mbox{\ensuremath{\,^{o}}C}$ and $400\mbox{\ensuremath{\,^{o}}C}$
increased respectively by $50\mbox{\%}$ and $300\%$. However, only
the latter were completely stable in time. We report fundamentally
different results for junctions fabricated using the RIE cleaning.
For sample $B$, after annealing at $200\mbox{\ensuremath{\,^{o}}C}$
we observe no change in the tunnel barrier resistance. This confirms
the stability of our junctions. After annealing at $400\mbox{\ensuremath{\,^{o}}C}$
we observe a decrease of the barrier by almost $50\mbox{\%}$. This
decrease is expected as the oxygen diffusion in the barrier starts
to be thermally activated for $T\gtrsim300\mbox{\ensuremath{\,^{o}}C}$\cite{GATES1984,Rodmacq2009}.

We have annealed at $200\mbox{\ensuremath{\,^{o}}C}$ the junctions
of a forth sample, $D$, similar to $C$, which was covered by a $PMMA$
resist layer. After the anneal, we observed a significant decrease
of $26\,\%$ of the tunnel resistance. This could be explained by
the contamination of the junction with hydrates$\left(-OH\right)$
from the resist, which combine with the oxide to form hydroxides (for
example $Al_{2}O_{3}(H_{2}O)$ or $Al_{2}O_{3}\cdot3(H_{2}O)$ ) and
thus decrease the barrier height, as suggested in Ref. \cite{GATES1984}.
These junctions were unstable and we measured a slow relaxation of
the barrier value towards the initial state. In one week, the barrier
increased by $3\mbox{\%}$. After re-annealing the sample $D$ under
a $PMMA$ resist layer at $200\mbox{\ensuremath{\,^{o}}C}$, the value
of the barrier returns to lower values. After 4 weeks, the barrier
relaxes again to a value $6\mbox{\%}$ larger. This reproducible increase-relaxation
process for the barrier value, after each anneal under a $PMMA$ resist
layer, can be explained as follows. During the annealing, the junction
absorbs hydrates which account for the barrier decrease. After the
lift-off of the resist, the slow desorption of hydrates during storage
accounts for the measured relaxation of the barrier. 

In conclusion, we have fabricated completely stable $Al/AlO_{x}/Al$
Josephson junctions, which were monitored for as long as 6 months,
without observing any changes in the barrier parameters. The crucial
processing step, allowing the fabrication of stable junctions, is
the optimized, isotropic RIE oxygen-plasma cleaning of the resist
residues, before the aluminum evaporation. We show that the junctions
are stable to annealing up to $200\mbox{\ensuremath{\,^{o}}C}$ in
vacuum. When annealed under a $PMMA$ resist layer, the junction barrier
decreases significantly and it also becomes unstable. Finally, measurements
on large ensembles of junctions at room temperature show that the
variance of the tunnel barrier height is in the range of only a few
percents. This spread is considered acceptable for most types of superconducting
circuits.

The samples have been fabricated in the \textit{Nanofab} platform
in Grenoble. The research has been supported by the European STREP
MIDAS and the French ANR QUANTJO.

\bibliographystyle{IEEEtran}

\begin{thebibliography}{10}
\bibitem{RIEparamGrenoble}
The cleaning process was conducted in a \textsc{P}lassys
  \textsc{MG}-200-\textsc{S} \textsc{RIE} equipment under an oxygen pressure of
  $3\cdot10^{-1}$ mbar and 10 \textsc{W}atts \textsc{RF} power, during 15
  seconds. The resist etching speed is $~1$nm/s.
\end{thebibliography}

\end{document}